\documentstyle[12pt,amssymb]{article}

\begin{document}
\tolerance=5000
\def\pp{{\, \mid \hskip -1.5mm =}}
\def\cL{{\cal L}}
\def\be{\begin{equation}}
\def\ee{\end{equation}}
\def\bea{\begin{eqnarray}}
\def\eea{\end{eqnarray}}
\def\tr{{\rm tr}\, }
\def\nn{\nonumber \\}
\def\e{{\rm e}}

\begin{titlepage}

\begin{center}
\Large {\bf Quantum escape of sudden future singularity}

\vspace*{5mm}

\normalsize

\large{ Shin'ichi Nojiri,$^\spadesuit$\footnote{Electronic mail: nojiri@nda.ac.jp,
snojiri@yukawa.kyoto-u.ac.jp}
and Sergei D. Odintsov$^{\clubsuit\heartsuit}$\footnote{Electronic mail:
odintsov@ieec.fcr.es Also at TSPU, Tomsk, Russia}}

\normalsize

\vfill

{\em $\spadesuit$ Department of Applied Physics,
National Defence Academy, \\
Hashirimizu Yokosuka 239-8686, Japan}
\medskip

{\em $\clubsuit$ Instituci\`o Catalana de Recerca i Estudis
Avan\c{c}ats (ICREA), \\Barcelona, Spain}
\medskip

{\em $\heartsuit$ Institut d'Estudis Espacials de Catalunya (IEEC), \\
Edifici Nexus, Gran Capit\`a 2-4, 08034 Barcelona, Spain}

\end{center}

\vfill

\begin{abstract}

We consider finite-time, future (sudden or Big Rip type) singularities
which may occur 
even when strong energy condition is not violated but equation of state
parameter is time-dependent. Recently, example of such singularity has
been presented by Barrow, we found another example of it.
Taking into account back reaction of conformal quantum fields near
singularity, it is shown explicitly that quantum effects may delay (or
make milder)
the singularity. It is argued that if the evolution to singularity is 
realistic, due to quantum effects the universe may end up in deSitter
phase before scale factor 
blows up.
This picture is generalized for braneworld where sudden singularity 
may occur on the brane with qualitatively similar conclusions.

\end{abstract}

\vfill


\end{titlepage}

\noindent
{\bf 1. Classical sudden singularities.}
The knowledge of the dark energy equation of state parameter $w$ is
extremely
important due to various reasons. According to current
astrophysical data this parameter is close (above or below) to $-1$.
If it less than $-1$ then such dark
energy (of phantom sort) drives the universe to finite time future
singularity (Big Rip)
\cite{brett, alexei, CKM} with catastrophic consequences for future
civilizations. According to various estimations in this case the universe
will exist
for several more billions of years at best, before ending up in a cosmic
doomsday. How realistic is such sudden future singularity? Is it common
property of only dark energy with $w$ less than $-1$?
The partial answer to second question is given in the recent paper
\cite{barrow} where it was shown that even for usual matter
 (and when strong energy condition is not violated!) the sudden future
singularity may occur.

In the present Letter we argue that near to future singularity the quantum
effects may become dominant and they can stop (or delay) the finite time
future singularity (for general discussion of classical singularities,
see book\cite{hawking}).
The same occurs for brane sudden singularities.
This indicates that  future Quantum Gravity era is quite possible and
deserves some attention.
In principle, the proof that sudden singularity does not occur maybe quite
simple. Indeed, one should write quantum-corrected FRW equations at late
time and show that there is stable non-singular solution (at some
conditions). For instance, such strategy in case of trace anomaly driven
inflation model shows that universe evolves to non-singular deSitter
space. We go even further, taking into account quantum effects
contribution as back-reaction near the future singularity. Even in such
complicated formulation, there are indications that finite time
singularity is soften (or delayed).

  Let us consider the
 generalization of Barrow's model \cite{barrow}. In this model,
 the matter has been
given implicitly via the FRW equations:
\be
\label{F1}
\rho=\frac{6}{\kappa^2}H^2\ ,\quad p=-\frac{2}{\kappa^2}\left(2\frac{dH}{dt} + 3H^2\right)\ .
\ee
The spatially-flat FRW metric is considered
\be
\label{dSP2}
ds^2= -dt^2 + a(t)^2\sum_{i=1,2,3} \left(dx^i\right)^2\ ,
\ee
and $H\equiv \frac{1}{a}\frac{da}{dt}$.
Eq.(\ref{F1}) shows that $\rho$ is always positive.

We now assume $H$ has the following form:
\be
\label{F2}
H(t)=\tilde H(t) + {A'}\left|t_s - t\right|^\alpha\ .
\ee
Here $\tilde H$ is a smooth, differentiable (infinite number of times, in
principle) function and ${A'}$ and $t_s$ are constants.
Another assumption is that a constant $\alpha$ is not a positive
integer. Then $H(t)$ has a singularity at $t=t_s$.
In case $\alpha$ is negative integer, the singularity is pole. Even if $\alpha$ is positive, in case $\alpha$ is not
an integer, there appears cut-like singularity.
 It is important for us that  singularity presents.
If we can consider the region $t>t_s$, there is no finite-time {\it
future}
singularity.
One also assumes the singularity belongs to the type
(\ref{F2}) although there are other types of singularities, say,
logarithmic ones.

When $\alpha>1$, one gets
\be
\label{F3}
\rho \sim -p \sim \frac{6}{\kappa^2}{\tilde H\left(t_s\right)}^2\ .
\ee
Hence $w=\frac{p}{\rho}=-1$, which may correspond to the positive
cosmological constant.

The case $0<\alpha<1$ corresponds to Barrow's model and when $t\sim t_s$, we find
\be
\label{F4}
\rho \sim \frac{6}{\kappa^2}{\tilde H\left(t_s\right)}^2\ ,\quad
p\sim \pm \frac{4{A'}\alpha}{\kappa^2}\left|t_s - t\right|^{\alpha-1}\ .
\ee
Here the plus sign in $\pm$ corresponds to $t<t_s$ case and the minus one to $t>t_s$.
In the following, the upper (lower) sign always corresponds to $t<t_s$ ($t>t_s$).
The  parameter of equation of state $w$ is given by
\be
\label{F5}
w=\pm \frac{2}{3}\frac{{A'}\alpha \left|t_s - t\right|^{\alpha-1}}{\tilde H\left(t_s\right)^2} \ .
\ee
Then it follows that $w$ is positive in two cases: one is ${A'}>0$ and
$t<t_s$, which directly corresponds
to Barrow's model,  and another is ${A'}<0$ $t>t_s$. In other cases, $w$ is negative.

When $-1<\alpha<0$, the energy density $\rho$ and the pressure is given by
\be
\label{F6}
\rho=\frac{6{A'}^2}{\kappa^2}\left|t_s - t\right|^{2\alpha}\ ,\quad
p\sim \pm \frac{4{A'}\alpha}{\kappa^2}\left|t_s - t\right|^{\alpha-1}\ .
\ee
The parameter of equation of state is
\be
\label{F7}
w=\pm \frac{2\alpha}{3{A'}}\left|t_s - t\right|^{-\alpha-1}\ ,
\ee
which diverges at $t=t_s$. Here $w$ is positive when ${A'}>0$ and $t<t_s$
or ${A'}<0$ and $t>t_s$.
The former case corresponds to future singularity even if $w$ is positive.
The singularity can be regarded as
a Big Rip. (For the recent comparison of phantom Big Rip with above type
of it, see \cite{ruth}. The mechanisms to escape of phantom Big Rip were
suggested in refs.\cite{phantom,emilio}.)

The case $\alpha=-1$ gives
\be
\label{F8}
\rho=\frac{6{A'}^2}{\kappa^2}\left|t_s - t\right|^{-2}\ ,\quad
p\sim -\frac{2}{\kappa^2}\left(\pm 2{A'} + 3{A'}^2\right)\left|t_s - t\right|^{-2}\ ,
\ee
which may correspond to the scalar field with exponential potential. The parameter $w$ is given by
\be
\label{F9}
w=-1\mp \frac{2}{{A'}}\ .
\ee
Near $t=t_s$, the universe is expanding if ${A'}>0$ and $t<t_s$ or ${A'}<0$ and $t>t_s$.
The former case corresponds to the phantom with $w<-1$. In the latter case,
 if $2>{A'}>0$, the equation of state describes
 the usual matter with positive $w$ and if ${A'}>2$, the matter may be the
quintessence with
$0>w>-1$.

If  $\alpha<-1$, one obtains
\be
\label{F9b}
\rho=-p=\frac{6{A'}^2}{\kappa^2}\left|t_s - t\right|^{2\alpha}\ ,
\ee
which gives $w=-1$ as for the cosmological constant case. In this case, however, there is rather
strong singularity at $t=t_s$.

We should note that $w$ is not constant in general. When $w$ is constant, except the case $w=-1$,
the energy density $\rho$ and the pressure $p$ are given by (\ref{F8}) and always behave as
$\left|t_s - t\right|^{-2}$. If we consider only expanding universe (not shrinking one), the future
singularity appears when $w<-1$, which corresponds to phantom. In the case
 with constant $w>-1$, there
does not appear the future singularity in the expanding universe although there is past or initial
singularity at $t=t_s$.

Notice that usually $w=-1$ corresponds to cosmological constant presence
but in general, in case that $w\to -1$ when $t\to t_s$
there can appear a singularity in the expanding universe as in (\ref{F3})
 and (\ref{F9b}). In case of (\ref{F3}),
the singularity is rather mild, and $H$ and $\frac{dH}{dt}$ (or $a$, $\frac{da}{dt}$, and $\frac{d^2 a}{dt^2}$)
are finite and $\frac{d^2 H}{dt^2}$ (or $\frac{d^3 a}{dt^3}$) diverges at
$t=t_s$. On the other hand, the case
(\ref{F9b}) shows very strong singularity, where the scale factor $a$ behaves as
$a\propto \e^{\mp \frac{{A'}}{\alpha +1}\left|t_s - t\right|^{\alpha+1}}$.

On the contrary, when
 $w$ is not constant but varies with $t$, even if the strong energy condition
\be
\label{B0}
\rho_m>0 \ ,\quad \rho_m + 3p_m>0
\ee
is satisfied, there can be a future singularity. An example is given in
\cite{barrow}, which corresponds
to (\ref{F4}). We have found another example given in (\ref{F6}) with ${A'}>0$ and $t<t_s$.

\noindent
{\bf 2. Quantum effects near Big Rip.}
When approaching the sudden future singularity,
the curvature invariants grow and typical energies increase. This
indicates that quantum effects
(or even quantum gravity effects) should become significant near future
singularity. They are usually associated with higher derivative terms.
Imagine that quantum effects of conformal fields are dominant,
as it occurs in some models of early universe, like in the
 trace anomaly driven inflation \cite{inflation}. Let us take
into account the conformal anomaly contribution as back-reaction near
singularity.
The correspondent  energy density $\rho_A$ and pressure $p_A$ are
(see, for instance, \cite{NOOfrw})
\bea
\label{hhrA3}
\rho_A&=&- 6 b'H^4 - \left({2 \over 3}b + b''\right)
\left\{ -6 H \frac{d^2 H}{dt^2} - 18 H^2 \frac{dH}{dt}
+ 3 \left(\frac{dH}{dt}\right)^2 \right\} \\
\label{hhrAA1}
p_A&=&b'\left\{ 6 H^4 + 8H^2 \frac{dH}{dt} \right\} \nn
&& + \left({2 \over 3}b + b''\right)\left\{ -2\frac{d^3 H}{dt^3} -12
H \frac{d^2 H}{dt^2} - 18 H^2 \frac{dH}{dt}  - 9 \left(\frac{dH}{dt}\right)^2 \right\}\ .
\eea
In general, with $N$ scalar, $N_{1/2}$ spinor, $N_1$ vector fields, $N_2$ ($=0$ or $1$)
gravitons and $N_{\rm HD}$ higher derivative conformal scalars, $b$, $b'$ and $b''$ are
given by
\bea
\label{bs}
&& b={N +6N_{1/2}+12N_1 + 611 N_2 - 8N_{\rm HD} \over 120(4\pi)^2}\nn
&& b'=-{N+11N_{1/2}+62N_1 + 1411 N_2 -28 N_{\rm HD} \over 360(4\pi)^2}\ , \quad b''=0\ .
\eea
We should note $b>0$ and $b'<0$ for the usual matter
 except the higher derivative conformal scalars.

Near the (future) singularity, the curvatures and their derivatives with
respect to $t$ become large.
Since the quantum correction includes the powers of the curvatures and
their $t$-derivatives,
the correction becomes large and important near the singularity.
It may occur that it delays (or stop) the singularity in the analogy
with trace anomaly driven inflation \cite{inflation}.

As the first example, Barrow's model \cite{barrow} is discussed.
Following \cite{barrow}, we first consider the case that the scale factor
$a(t)$ is given by
\be
\label{B1}
a(t)=A + B t^q + C\left(t_s - t\right)^n\ .
\ee
Here $A>0$, $B>0$, $q>0$, and $t_s>0$ are constants and $C=-At_s^{-n}$.
 One also assumes
$t<t_s$ and $2>n>1$. There appears a singularity when $t\to t_s$, where $\frac{1}{a}
\frac{d^2 a}{dt^2}\to + \infty$.
Since
\be
\label{CQ1}
H=\frac{qB t^{q-1} - Cn\left(t_s - t\right)^{n-1}}{A+ B t^q + C\left(t_s - t\right)^n}
\sim \frac{qB t_s^{q-1}} {A+ B t_s^q} - \frac{Cn\left(t_s - t\right)^{n-1}}{A+ B t_s^q}\ ,
\ee
when $t\sim t_s$,  the parameters in (\ref{F2}) correspond to
\be
\label{CQ2}
A'=- \frac{Cn}{A+ B t_s^q}\ ,\quad \alpha=n-1\ .
\ee
The classical FRW equations
\be
\label{B2}
{6 \over \kappa^2}H^2 =\rho_m\ ,\quad \frac{2}{\kappa^2 a}\frac{d^2 a}{dt^2}
= - \frac{\rho_m + 3p_m}{6}\ ,
\ee
show that
\be
\label{B3}
\rho_m \sim \frac{6q^2 B^2 t_s^{2q-2} }{\kappa^2 \left(A + B t_s^q\right)}>0\ ,\quad
p_m \sim -\frac{Cn(n-1)\left(t_s - t\right)^{n-2}}{A+B t_s^q}>0\ ,
\ee
near the singularity $t\sim t_s$.
The energy density $\rho_m$ is finite but the pressure $p_m$ diverges.
Singularity presents but the strong energy condition (\ref{B0}) is
satisfied since
$\rho$ and $p$ are positive.

With the account of the quantum corrections  (\ref{hhrA3}) and
(\ref{hhrAA1}), the
classical FRW equations (\ref{B2}) are modified as
\be
\label{B4}
{6 \over \kappa^2}H^2 =\rho_m + \rho_A \ ,\quad \frac{2}{\kappa^2 a}\frac{d^2 a}{dt^2}
= - \frac{\rho_m + \rho_A + 3p_m + 3p_A}{6}\ .
\ee
As clear from the expressions  (\ref{hhrA3}) and (\ref{hhrAA1}),
 $\rho_A$ and $p_A$ become singular at $t=t_s$.
 It follows from (\ref{B4})
\bea
\label{B5}
\rho_m &\sim& - \rho_A \nn
&\sim& 6\left(\frac{2}{3}b + b''\right)
\frac{BC q n(n-1)(n-2) t_s^{q-1} \left(t_s - t\right)^{n-3}}{\left(A + B t_s^q\right)^2}
>0\ ,\nn
p_m &\sim& -p_A \nn
&\sim& 2\left(\frac{2}{3}b + b''\right)
\frac{C n(n-1)(n-2)(n-3) \left(t_s - t\right)^{n-4}}{A + B t_s^q}<0\ .
\eea
Thus, the energy density $\rho_m$ is positive but diverges at $t=t_s$.
Since the pressure $p_m$ is negative and more singular than $\rho_m$,
the strong energy condition (\ref{B0}) is violated and  also  $\rho + p<0$.

For the example (\ref{B3}), the singularity at $t=t_s$ might become milder
due to the quantum correction. To see this, we assume $4>n>3$, instead of $2>n>1$, in
(\ref{B1}). For this choice $a$, $\frac{da}{dt}$, $\frac{d^2a}{dt^2}$,
and $\frac{d^3 a}{dt^3}$ are finite when $t=t_s$ but $\frac{d^4a}{dt^4}$ diverges
there. Then the singularity is rather milder than the case $2>n>2$, where $\frac{d^2 a}{dt^2}$ diverges.
Using (\ref{B4}) around $t=t_s$, one gets
\bea
\label{B6}
\rho_m &=& {6 \over \kappa^2}H^2 - \rho_A \nn
&\sim & \frac{6q^2 B^2 t_s^{2q-2} }{\kappa^2 \left(A + B t_s^q\right)}
 -\frac{6b'B^4 q^4 t_s^{4q-4}}{\left(A+Bt_s^q\right)^4} \nn
&& - 3\left(\frac{2}{3}b + b''\right)\left(
 -\frac{B^2 q^2(q-1)(q-3) t_s^{2q-4}}{\left(A+Bt_s^q\right)^2}
 -\frac{2B^3 q^3(q-1) t_s^{3q-4}}{\left(A+Bt_s^q\right)^3} \right. \nn
&& \left. + \frac{3B^4 q^4 t_s^{4q-4}}{\left(A+Bt_s^q\right)^4}
\right) \ ,\nn
p_m &\sim& -p_A \nn
&\sim& 2\left(\frac{2}{3}b + b''\right)
\frac{C n(n-1)(n-2)(n-3) \left(t_s - t\right)^{n-4}}{A + B t_s^q}\ .
\eea
In the above expression, $\rho_m$ is not positive in general.
However,
 in a special
case  $b=-b'>0$, $q=1$, it reduces to
\be
\label{B7}
\rho_m \sim \frac{6 B^2}{\kappa^2 \left(A + B t_s\right)}\ ,
\ee
which is positive.  $p_m$ in (\ref{B6}) is,
however, negative since $C=-At_s^{-n}<0$ in \cite{barrow}. In principle,
 there is no any strong reason to exclude the case that $C$ is
positive. In the case $C>0$, if one replaces
\bea
\label{B8}
2\left(\frac{2}{3}b + b''\right)C n(n-1)(n-2)(n-3)
&\to& - C n(n-1)\ ,\nn
n &\to& n+2\ ,
\eea
in  $p_m$ (\ref{B6}), we obtain  $p_m$  (\ref{B3}).
The arguments may be reversed. If we start with the matter  (\ref{B3})
(say, with $q=1$), due to the quantum correction, we obtain the following scale
factor $a$ (instead of (\ref{B1})):
\be
\label{B9}
a(t)=A + B t + C'\left(t_s - t\right)^{n'}\ .
\ee
Here $A>0$, $B>0$, $C>0$, and $t_s>0$ are constants and $4>n'>3$.
This demonstrates that the singularity becomes much milder due to the
quantum correction.

In general, near the singularity $\rho_A$ becomes dominant. The natural
assumption is then
\be
\label{CQA1}
\rho\sim -\rho_A\ .
\ee
For the case given in (\ref{F6}), we find
\be
\label{CQA2}
H(t)={\tilde H}'(t) + A'\left|t_s - t\right|^{2\alpha+1}\ ,
\ee
Here ${\tilde H}'(t)$ is a smooth and differentiable function and $A'$
satisfies the following
equation:
\be
\label{CQA3}
\frac{A^2}{\kappa^2}=-2\left(\frac{2}{3}b + b''\right){\tilde H}'\left(t_s\right)
\left(2\alpha+1\right)\alpha A' >0\ .
\ee
For the matter corresponding to (\ref{F6}), $H$ behaves as
 $H\sim \left|t_s - t\right|^{-1 \sim 0}$ but with the account of
 the quantum correction, one obtains $H\sim \left|t_s - t\right|^{-1 \sim
1}$. Thus, the singularity
becomes milder under quite general conditions. One should not forget that
account of quantum gravity
corrections (see recent example in \cite{emilio}) may improve the
situation with the singularity occurence even more.
Finally, the quantum effective action which is
usually  non-local and higher-derivative typically leads to violation of
energy conditions on quantum level. Hence, in many cases it may support
the singularities. Nevertheless, it is interesting that like in anomaly
driven inflation there occur the stable regimes where quantum effects may
drive (at least, partially) the future universe out of the singularity.
Moreover, like in anomaly-driven inflation (with correspondent boundary
conditions) one may argue that even future deSitter universe is quite
realistic possibility as the corresponding solution of quantum-corrected
FRW equation exists.

\noindent
{\bf 3. Quantum escape in the brane-world.}
It is interesting that the above consideration of sudden future
singularity may be easily generalized
for brane-world models (for recent review, see \cite{roy,review,NOOfrw}
and refs therein).
Let the 3-brane is embedded into the 5d bulk space as in \cite{SMS}.
Let $g_{\mu\nu}$ be the metric tensor of the bulk space and $n_\mu$ be the unit vector
normal to the 3-brane. The metric $q_{\mu\nu}$ induced on the brane has the
following form:
\be
\label{S1b}
q_{\mu\nu}=g_{\mu\nu} - n_\mu n_\nu\ .
\ee
The initial action is
\be
\label{S00}
S=\int d^5 x \sqrt{-g}\left\{ {1 \over \kappa_5^2} R^{(5)} - 2\Lambda + \cdots \right\}
+ S_{\rm brane}(q)\ .
\ee
Here,  the 5d quantities are denoted by the suffix $^(5)$ and 4d ones by
$^(4)$. In (\ref{S00}), $\cdots$ expresses the matter  contribution
and $S_{\rm brane}$ is the action on the brane, which will be specified later.
The bulk Einstein equation is given by
\be
\label{S2bb}
{1 \over \kappa_5^2}\left( R^{(5)}_{\mu\nu} - {1 \over 2}g_{\mu\nu} R^{(5)}\right)
= T_{\mu\nu}\
\ee
If one chooses the metric near the brane as:
\be
\label{S2b}
ds^2 = d\chi^2 + q_{\mu\nu} dx^\mu dx^\nu\ ,
\ee
the energy-momentum tensor $T_{\mu\nu}$ has the following form:
\be
\label{S2c}
T_{\mu\nu} = T_{\mu\nu}^{\rm bulk\ matter} - \Lambda g_{\mu\nu}
+ \delta(\chi)\left(-\lambda q_{\mu\nu} + \tau_{\mu\nu}\right)\ .
\ee
Here $T_{\mu\nu}^{\rm bulk\ matter}$ is the energy-momentum tensor of the bulk matter,
$\Lambda$ is the bulk cosmological constant, $\lambda$ is the tension of the brane,
and $\tau_{\mu\nu}$ expresses the contribution due to brane matter.
Without the bulk matter ($T_{\mu\nu}^{\rm bulk\ matter}=0$),
 following the procedure in \cite{SMS}, the bulk Einstein equation can be mapped into
the equation on the brane:
\bea
\label{S2d}
&& {1 \over \kappa_5^2} \left( R^{(4)}_{\mu\nu} - {1 \over 2} q_{\mu\nu}R^{(4)}\right) \nn
&& = - {1 \over 2}\left( \Lambda + {\kappa_5^2 \lambda^2 \over 6} \right) q_{\mu\nu}
+ {\kappa_5^2 \lambda \over 6}\tau_{\mu\nu}
+ \kappa_5^2\pi_{\mu\nu} - {1 \over \kappa_5^2}E_{\mu\nu}\ .
\eea
Here $\pi_{\mu\nu}$ is given by
\be
\label{S2e}
\pi_{\mu\nu}=-{1 \over 4}\tau_{\mu\alpha}\tau_\nu^{\ \alpha} + {1 \over 12}\tau \tau_{\mu\nu}
+ {1 \over 8}q_{\mu\nu}\tau_{\alpha\beta} \tau^{\alpha\beta} - {1 \over 24}q_{\mu\nu}\tau^2\ .
\ee
On the other hand, $E_{\mu\nu}$ is defined by the bulk Weyl tensor $C^{(5)}_{\mu\nu\rho\sigma}$
but as the bulk spacetime is anti-deSitter one, where the Weyl tensor
vanishes,
we put  $E_{\mu\nu}=0$. Like in \cite{RS} the tension of the brane
$\lambda$ is fine-tuned
to be
\be
\label{BBB1}
0=\Lambda + {\kappa_5^2 \lambda^2 \over 6}\ .
\ee
The metric on the brane is supposed to be the FRW metric (\ref{dSP2}).
Let the energy density $\rho_m$ and the pressure $p_m$ of the matter on
the
brane are given as in (\ref{B3}):
\be
\label{BBB2}
\tau_{tt}=\frac{\rho_m}{2}\ ,\quad \tau_{ij}={p_m \over 2}a^2\delta_{ij}\ ,\quad
\rho_m\sim \rho_0\ ,\quad p_m\sim p_0\left(t_s - t\right)^{-\epsilon}\ .
\ee
Here $\rho_0>0$, $p_0>0$, and $1>\epsilon>0$ are constants. Then the strong energy condition (\ref{B0})
is satisfied.
Now the effective Einstein equation (\ref{S2d}) has the following form:
\be
\label{BBB3}
\frac{6}{\kappa^2}H^2=\rho_m + {\kappa_5^2 \over 24}\rho_m^2\ ,\quad
\frac{2}{\kappa^2 a}\frac{d^2 a}{dt^2}=-\frac{\rho_m + 3p_m}{6} + \frac{\kappa_5^2}{72}\left(3p_m \rho_m
+ 2\rho_m^2\right)\ .
\ee
Here
\be
\label{BBB4}
\kappa^2\equiv \frac{\lambda\kappa_5^2}{6}\ .
\ee
When $t\sim t_s$, one gets
\bea
\label{BBB4b}
&& H \sim \sqrt{\frac{\kappa^2}{6}\left(\rho_0 + {\kappa_5^2 \over 24}\rho_0^2\right)}\ ,\nn
&& \frac{2}{\kappa^2 a}\frac{d^2 a}{dt^2}\sim -\frac{p_m}{2} + \frac{\kappa_5^2}{24}p_m 
\sim -\frac{p_0}{2}\left(1+\frac{\kappa_5^2\rho_0}{12}\right)
\left(t_s - t\right)^{-\epsilon}\ ,
\eea
which shows quantatively the same behavior with that given by (\ref{B1})
with the following identification
\be
\label{BBB5}
\rho_0= \frac{6q^2 B^2 t_s^{2q-2} }{\kappa^2 \left(A + B t_s^q\right)}\ ,\quad
p_0 = -\frac{Cn(n-1)}{A+B t_s^q}\ ,\quad \epsilon=2-n\ .
\ee
Hence, even for the brane-world, the qualitative behavior of the
singularity is not changed (for earlier discussion of finite time singularities in brane-world, see \cite{varun}).

The contribution from the conformal anomaly as a quantum correction from
the matter on the
brane may be again included. This can be done by replacing $\rho_m$ and
$p_m$ in (\ref{BBB3}) with $\rho_m + \rho_A$ and
$p_m + p_A$, respectively. Near singularity $t\sim t_s$, one gets
\be
\label{BBB6}
p_m\sim -p_A\ ,
\ee
that is,
\be
\label{BBB7}
p_0\left(t_s - t\right)^{-\epsilon}\sim 2\left(\frac{2}{3}b + b''\right)\frac{d^3 H}{dt^3}\ ,
\ee
which gives
\be
\label{BBB8}
H=\left(\frac{2}{3}b + b''\right)\frac{p_0 \left(t_s - t\right)^{3-\epsilon}}{2(1-\epsilon)(2-\epsilon)(3-\epsilon)}
+ \tilde H(t)\ .
\ee
Here $\tilde H(t)$ is a smooth, differentiable (at least, three times)
function. Hence, the qualitative behavior is
similar to that given by (\ref{B9}) with identification
\be
\label{BBB9}
\epsilon=4-n'\ .
\ee
Thus, the singularity becomes much milder.
The quantum effects of conformal matter help to realize the scenario of
brane inflation
called Brane New World \cite{NOZ,HHR,NO}.
As it follows from above consideration, the account of such quantum
effects  may help to avoid (or, at least to delay) the future sudden
singularity. Moreover, as anomaly driven brane inflationary solution
of FRW equations exists \cite{NOZ,HHR,NO}, one can argue that at similar
conditions the future brane universe ends up in deSitter phase.

The final picture (if realistic) looks quite strange. Quantum gravity
era of the early universe helps in the realization of inflation.
Subsequently, after quite complicated history the universe evolves to dark
energy dominated phase which eventually may evolve to sudden singularity.
The approach to finite-time future singularity brings back to life the
Quantum Gravity era. In its turn, quantum effects may stop further
evolution into the singularity what decreases the typical energies.
In resulting, non-singular universe the quantum effects become negligible.
Eventually, new circle of inflation (deflation) starts.
This all reminds very much oscillating universe.

\noindent
{\bf Acknowledgments}
We thank J. Lidsey, A. Starobinsky and P. Townsend for related
discussions.
This investigation has been supported in part by the
Ministry of Education, Science, Sports and Culture of Japan under
grant n.13135208 (S.N.), by the RFBR grant 03-01-00105, LRSS
grant 1252.2003.2 and BFM2003-00620 grant (S.D.O.).

\end{document}